\pdfoutput=1
\documentclass[aps,twocolumn,floats,floatfix,superscriptaddress]{revtex4}
\usepackage{graphicx,amssymb,amsmath,subfigure,cancel}
\usepackage[utf8]{inputenc}
\usepackage{color}
\definecolor{red}{rgb}{1,0,0}

\definecolor{blue}{rgb}{0,0,1}

\begin{document}

\title{{Successful} network inference from time-series data {using} Mutual Information Rate}

\author{E. Bianco-Martinez}
\affiliation{Institute for Complex Systems and Mathematical Biology, SUPA, University of Aberdeen, King's College, AB24 3UE Aberdeen, UK}

\author{N. Rubido}
\affiliation{Instituto de F\'{i}sica, Facultad de Ciencias, Universidad de la Rep\'{u}blica, Igu\'{a} 4225, Montevideo, 11200, Uruguay}
\affiliation{Institute for Complex Systems and Mathematical Biology, SUPA, University of Aberdeen, King's College, AB24 3UE Aberdeen, UK}

\author{Ch. G. Antonopoulos}
\affiliation{Department of Mathematical Sciences, University of Essex, Wivenhoe Park, CO4 3SQ Colchester, UK}

\author{M. S. Baptista}
\email{murilo.baptista@abdn.ac.uk}
\affiliation{Institute for Complex Systems and Mathematical Biology, SUPA University of Aberdeen, King's College, AB24 3UE Aberdeen, UK}
 
\date{\today}

\begin{abstract}
{This work uses an} information-based methodology to infer the connectivity of complex systems {from observed} time-series data. We first derive {analytically an expression for the Mutual Information Rate (MIR), namely, the amount of information exchanged per unit of time, that can be used to estimate the MIR between two finite-length low-resolution noisy time-series}, and then apply it {after a proper normalization} for the identification of the connectivity structure of {small networks of interacting dynamical systems. In particular, we show} that our methodology successfully infers the {connectivity} for heterogeneous networks, different time-series lengths {or coupling strengths, and even} in the presence of {additive} noise. Finally, we show that our methodology based on MIR successfully infers the connectivity of networks composed of nodes with different time-scale dynamics, where inference based on Mutual Information fails.
\end{abstract}


\maketitle

\textbf{The Mutual Information Rate (MIR) measures the time rate of information exchanged between two non-random and correlated variables. Since variables in complex systems are not purely random, MIR is {an appropriate} quantity to access the amount of information {exchanged} in complex systems. However, its calculation requires infinitely long {measurements with} arbitrary resolution. Having in mind that it is impossible to perform infinitely long measurements with {perfect} accuracy, this work shows how to estimate MIR taking into consideration this fundamental limitation {and how to use it} for the characterization and understanding of dynamical and complex systems. {Moreover, we} introduce a {novel} normalized form of MIR that successfully infers the structure of {small networks of interacting dynamical systems}. The proposed inference methodology {is} robust in the presence of additive noise, different time-series lengths, and heterogeneous node dynamics and coupling strengths. {Moreover,} it also outperforms {inference methods based on Mutual Information when analysing} networks formed by nodes possessing different time-scales.}\vspace{-1pc}

\section{Introduction}
We understand a complex system as a system with a large number of interacting components whose aggregated behaviour is non-linear and undetermined from the behaviour of the individual components {\cite{Barrat2008}}. If we now consider these components as nodes of a network, and the underlying physical interaction between any two nodes as links, one way to understand these complex systems is by studying its topological structure, namely, the network connectivity. In natural complex systems, the connectivity of the components is often unknown or is difficult to detect by physical methods due to {large system-sizes}. Hence, it is of interest to infer the network structure that represents the physical interaction between time-series collected from the dynamics of the nodes.

Although network inference in non-linear systems has been extensively studied in recent years {using Cross-Correlation or Mutual Information \cite{todos,Tirabassi2015,butte}, recurrences \cite{mamen,mamen2010,Hempel2011}, functional dynamics \cite{ta,Timme2007,Timme2011,casadiego}, and Granger Causality \cite{granger,Feng2010,Bressler2010}, to name a few}, it still presents open challenges. The fundamental reason is that non-linearities, even in the absence of noise, produce behaviour that hinders the correct identification of {existing or non-existing} underlying direct physical dependence between {any} pair of nodes.

In this paper, we introduce an information-based methodology to infer the structure of complex systems from time-series data. Our methodology is based on a normalized form of {an estimated} Mutual Information Rate (MIR), the rate by which information is exchanged per unit of time between any two components. MIR is an appropriate measure to quantify the exchange of information in systems with correlation { \cite{palus,schreiber,
Baptistaetal2012}. In particular,} authors in Ref.~\cite{Baptistaetal2012} show how to calculate MIR in the case a Markov partition is attainable, which is generally {extremely difficult to find or unknown. Here,} we {first} show how MIR can be approximately calculated for time-series data of finite length and low-resolution. {Then, we propose a normalization of the estimated MIR that allows for a successful inference about the dependence structure of} {small networks of interacting dynamical systems}, when Markov partitions are unknown. Our findings show {that the estimated normalized} MIR {allows for a {successful} inference of the structure of small networks} {even in the presence of additive noise, parameter heterogeneities {and} different coupling strenghts.} {Moreover, our normalized estimated MIR outperforms the use of Mutual Information (MI) based} {inference} {when different time-scale dynamics are present in the networks}.

The paper is organized as follows. In Sec.~\ref{section_methods_and_material}, we introduce two information-based measures, the MI and the MIR. We discuss the theoretical aspects of their definitions and show how they are related to each other. In Sec.~\ref{sec:section_models}, we introduce the models used to create the complex system dynamics studied in this work. In Sec.~\ref{sec:methodology}, we explain our methodology to calculate an approximation {value} of MIR and introduce the normalized MIR. Section~\ref{sec:results} shows how we apply our methodology to different coupled maps and to a neural network in which the dynamics of the nodes is described by the Hindmarsh-Rose neuron model \cite{Hindmarsh-Rose_original_paper,Hindmarsh-Rose_Baptista}. Finally, in Sec.~\ref{sec:conclusions} we discuss our work and discuss our findings.

\section{Background}\label{section_methods_and_material}
Information can be produced in a system and it can be transferred between its different components {\cite{Baptistaetal2012,palus,schreiber,
Antonopoulosetal2015,Machiorietal2012,Mandaletal2013,Antonopoulosetal2014}}. If transferred, at least two components that are physically interacting by direct or indirect links should be involved. In general, these components can be time-series, modes, or related functions of them, defined on subspaces or projections of the state space of the system. In this work, we study the amount of information transferred per {unit of time}, {i.e., the Mutual Information Rate (MIR)}, between any two components of a {system,} to determine if a link between them {exists. The existence of a} link between two units means there is a bidirectional connection between them {due to their interaction}.

\subsection{Mutual Information}\label{subsection_MI}
The Mutual Information (MI) \cite{shannon} between two random variables, $X$ and $Y$, of a system is the amount of uncertainty one has about $X$ ($Y$) after observing $Y$ ($X$). {Specifically,} {MI} is given by  \cite{shannon,kullback1959,dobrushin}
\begin{equation}\label{eq:Shannon_MI}
I_{XY}(N) = H_{X} + H_{Y} - H_{XY},
\end{equation}
where $H_{X} =-\sum_{i=1}^N P_{X}(i)\log \left(P_{X}(i)\right)$ and $H_{Y}=-\sum_{j=1}^N P_{Y}(j)\log \left(P_{Y}(j)\right)$ are the marginal entropies of $X$ and $Y$ (Shannon entropies) respectively, and $H_{XY}=-\sum^{N^2}_{i,j=1}P_{XY}(i,j)\log \left(P_{XY}(i,j)\right)$ is the joint entropy between $X$ and $Y$. $P_{X}(i)$ is the probability of a random event $i$ to happen in $X$, $P_{Y}(j)$ is the probability of a random event $j$ {to} happen in $Y$, and $P_{X,Y}(i,j)$ is the joint probability of events $i$ and $j$ to occur simultaneously in variables $X$ and $Y$. $N$ is the number of random events in both variables, $X$ and $Y$. 

{In particular, Eq.~}\eqref{eq:Shannon_MI} can be written equivalently as
\begin{equation}\label{MI_definition_probabilities}
I_{XY}(N)=\sum^N_{i}\sum^N_{j}P_{XY}(i,j)\log\left(\frac{P_{XY}(i,j)}{P_{X}(i)P_{Y}(j)}\right).
\end{equation}
This equation can be interpreted as the strength of the dependence between two random variables $X$ and $Y$ \cite{kullback1959}. When $I_{XY} = 0$, the dependence strength between $X$ and $Y$ is null, consequently, $X$ and $Y$ are independent.

The computation of $I_{XY}(N)$ from time-series is a subtle task. Firstly, it requires the calculation of probabilities computed on an appropriate probabilistic space {on which} a partition can be defined. Secondly, {$I_{XY}(N)$} is a measure suitable for the comparison between pairs of components of the same system but not between different systems. {The reason is that different systems can have} different correlation {decay times}, {\cite{Eckmann1985,Baptistaetal2008B,murilo2010} hence, different characteristic time-scales}.

There are three main approaches to compute MI, and the variation resides in the different ways to compute the probabilities involved in Eq.~\eqref{MI_definition_probabilities}. The first one is the bin or histogram method, which finds a suitable partition of the 2D space on equal or adaptive-size cells \citep{darbellay,fraser}. The second one employs density kernels, where a kernel estimation of the probability density function is used \citep{kernel1,kernel}. The last one computes MI by estimating probabilities from the distances between closest neighbours \cite{krakov}. In this work, we adopt the first method and compute probabilities in a partition of equally-sized cells in the probabilistic space generated by two variables $X$ and $Y$. It is well known that this approach, proposed in \cite{butte} and studied in \cite{steuer}, overestimates the value of $I_{XY}(N)$ for random systems or non-Markovian partitions \cite{steuer,herzel}. In particular, the authors explain two basic reasons for the overestimation of MI. The finite resolution of a non-Markovian partition and the finite length of the recorded time-series. According to \cite{steuer,herzel}, these errors are systematic and are always present in the computation of MI for an arbitrary non-Markovian partitions. {Here, we avoid these systematic error by creating a novel normalization when dealing with the MIR.}

For the numerical computation of $I_{XY}(N)$ [Eq.~\eqref{MI_definition_probabilities}], we use the approach reported in Refs.~\cite{butte,Baptistaetal2012}. We define a probabilistic space $\Omega${, where} $\Omega$ is formed by the time-series data {observed from a pair of nodes,} $X$ and $Y$, of a complex system. Then, we partition $\Omega$ into a grid of $N \times N$ fixed-sized cells. The length-side of each cell, $\epsilon$, is then {set to} $\epsilon = 1/N$. Consequently, the probability of having an event $i$ for variable $X$, $P_{X}(i)$, is the fraction of points found in row $i$ of the partition $\Omega$. Similarly, $P_{Y}(j)$ is the fraction of points that are found in column $j$ of $\Omega$, and $P_{XY}(i,j)$ is the joint probability  computed from the fraction of points that are found in cell $(i,j)$ of the same partition, where $i,\;j=1,\ldots,N$. We emphasize here that $I_{XY}(N)$ depends on the partition considered for its calculation as $P_{X}$, $P_{Y}$, and $P_{XY}$ attain different values for different cell-sizes $\epsilon$.

\subsection{Mutual Information Rate}\label{subsection_MIR}
Due to the issues arising from the definition of MI in terms of its partition dependence, the authors in Ref.~\cite{Baptistaetal2012} have demonstrated how to calculate {the MIR} for two time-series of finite length {irrespective of the partitions, instead of using the MI}. This quantity is invariant with respect to the resolution of the partition \cite{Baptistaetal2012}. {In particular, and} for infinitely long time-series, MIR is theoretically defined as the mutual information exchanged per {unit of time} between $X$ and $Y$ \cite{shannon,dobrushin,black}. Specifically,
\begin{eqnarray}
\nonumber
\mbox{MIR}_{XY}=\lim_{N\rightarrow \infty} \lim_{L\rightarrow \infty}\sum_{i=1}^{{L-1}}\frac{I_{XY}(i+1,N)-I_{XY}(i,N)}{L} \\ \nonumber
=\lim_{N\rightarrow \infty} \lim_{L\rightarrow \infty}\frac{I_{XY}({{L}},N)-I_{XY}(1,N)}{L} \\
=\lim_{N\rightarrow \infty} \lim_{L\rightarrow \infty}\frac{I_{XY}(L,N)}{L},\label{MIR_limits}
\end{eqnarray}
where $I_{XY}(L,N)$ represents the MI of Eq.~\eqref{eq:Shannon_MI} between random variables $X$ and $Y$, considering trajectories of length $L$ that follow an itinerary over boxes in a grid with an infinite number of cells $N$. Since $I_{XY}$ is a symmetric function with respect to $X$ and $Y$, $\mbox{MIR}_{XY}=\mbox{MIR}_{YX}$. We also note that the term $\frac{I_{XY}(1,N)}{L}$ tends to zero in the limit of infinitely long trajectories, $L\to \infty$.

The authors in Ref.~\cite{Baptistaetal2012} show that if a partition with $N$ cells is a Markov partition of order $T$, then MIR can be {estimated from finite-length and low-resolution time-series} (since the limits in Eq.~\eqref{MIR_limits} {are not necessary}) by using 
\begin{equation}\label{MIR_definition_epsilon}
\mbox{MIR}_{XY} = \frac{I_{XY}(N)}{T(N)},
\end{equation}
\noindent
{where both $T(N)$ and $N$ are finite quantities. Notice that an order $T$ partition can only generate statistically significantly probabilities if there is in each cell a sufficiently large amount of points (see Eq. (\ref{noc})). Besides, points in a cell must spread over the probabiliistic space $\Omega$ after $T$ iterations. So, the length of the time-series must be reasonably larger than $T$.} 

 {In Sec. \ref{sec:methodology}, we make a novel demonstration of Eq. (\ref{MIR_definition_epsilon}), from which it becomes clear why MIR can be estimated from finite-length and low-resolution time-series.} In this equation, $I_{XY}(N)$ is the MI between $X$ and $Y$, considering probabilities that are calculated in a Markov partition, and $T(N)$ represents the shortest time for the correlation between $X$ and $Y$ to be lost for that particular Markov partition. $T(N)$ also represents the time after which the evolution of a chaotic system is unpredictable. Moreover, this time is of the order of the shortest Poincar{\'e} return-time \cite{murilo2010} and {is} related to the order $O$ Markov partition, where $O$ indicates that the future state of a random variable $X$ is independent on its $(O-1)$s previous states and is independent on the states of $X$ for an order $O_{-T}$. 

\section{Models for our complex systems} \label{sec:section_models}
We adopt various topologies {for the networks} and {various} dynamics {for the components} {of the complex systems considered}. Hence, the network inference, which represents {the detection of} the topological structure of the component's interactions, is done from the time-series that are recorded for each component. In particular, we divide the analysis {on discrete} {and on continues time-series components}.

\subsection{Networks with discrete-time units} \label{sec:discrete_models}
The dynamics of the class of discrete complex systems that are of interest here are described by the following equation \cite{kaneko} 
\begin{equation} \label{kaneko}
x^{i}_{n+1}=f(x^{i}_{n},r)(1-\alpha)+\frac{\alpha}{k_{i}}\sum_{j=1}^M \boldsymbol{A}_{ij}f(x^{j}_{n},r),
\end{equation}
where $x^{i}_n$ is the $n$-th iterate of map $i$, where $i=1,\ldots,M$ and $M$ is the number of maps (nodes) of the system, $\alpha\in[0,1]$ is the coupling strength, $\boldsymbol{A}_{ij}$ is the binary adjacency matrix (with entries $1$ or $0$, depending on whether there is a connection between nodes $i$ and $j$ or not, respectively) that defines the structural connectivity in the network, $r$ is the dynamical parameter of each map, $k_{i}=\sum_{j=1}^M \boldsymbol{A}_{ij}$ is the node-degree, and $f(x_{n},r)$ is the considered map. Particularly, we use
\begin{eqnarray}
f(x_{n},r)&=&rx_{n}(1-x_{n}),\;\; \text{and} \label{logistic_map_eq}\\
f(x_{n},r)&=&x_{n}+r-\frac{K}{2\pi}\sin(2\pi x_{n})\mod 1 \label{circle_map_eq}.
\end{eqnarray}
For the logistic map \cite{May1976} {of Eq.~\eqref{logistic_map_eq}}, we use $r=4$ (if it is not explicitly mentioned), that corresponds to fully developed chaos, whereas for the circle map \cite{Colletetal1980} {of Eq.~\eqref{circle_map_eq}} we use $r=0.35$ and $K\approx6.9115$, following Ref.~\cite{todos}, for the same reason.

\begin{figure}[htbp]
\centering
\includegraphics[scale=0.3]{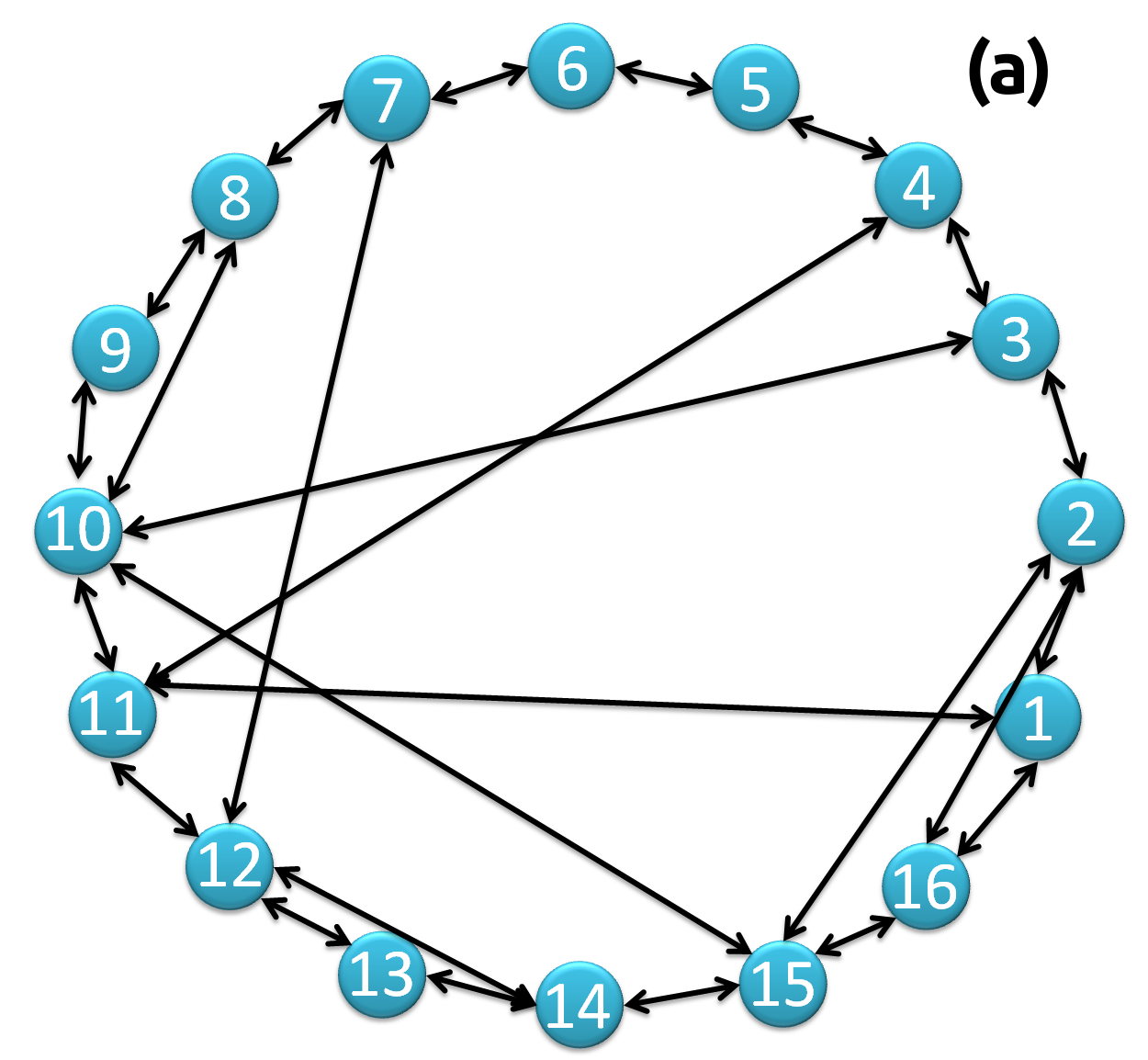}
\includegraphics[scale=0.2]{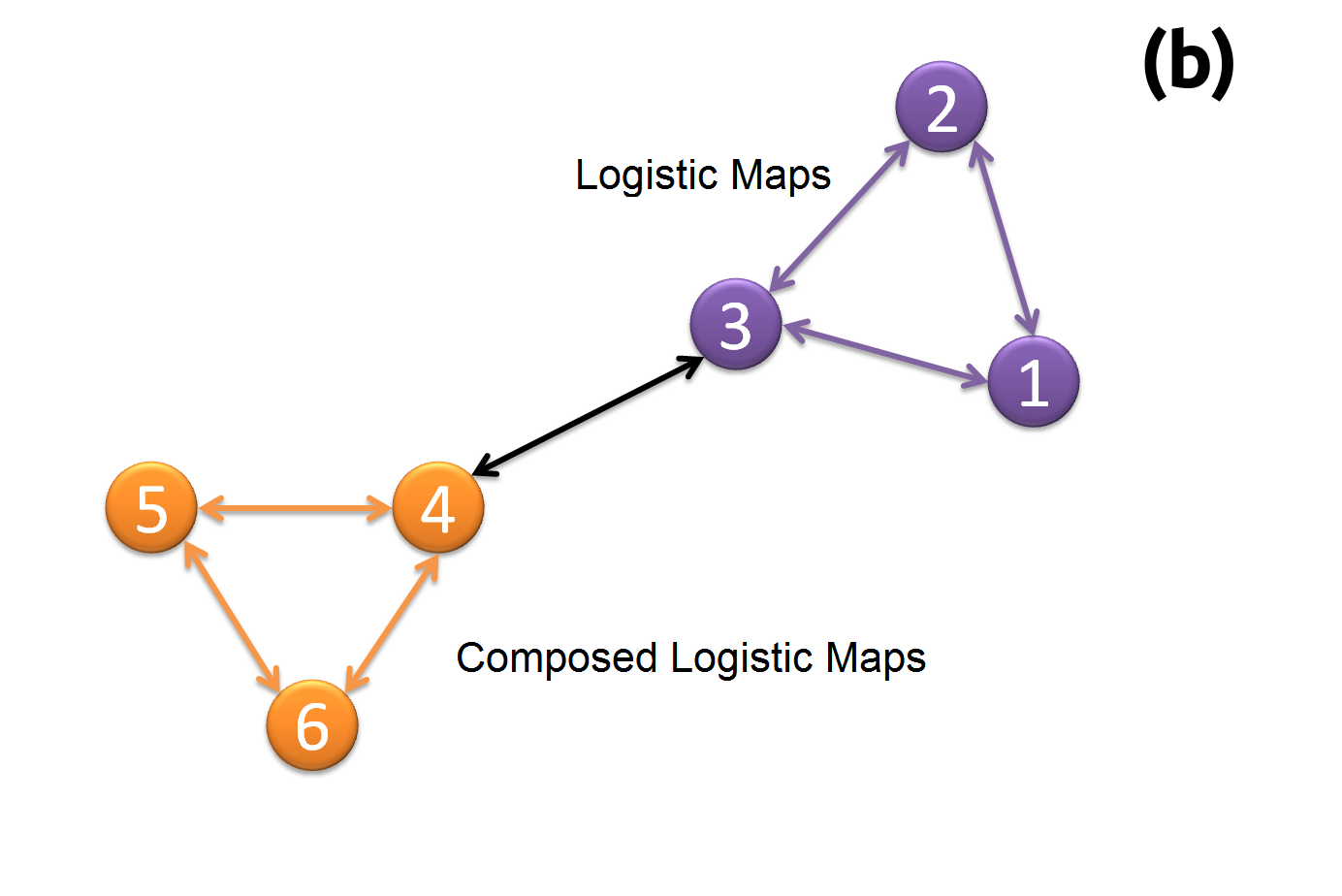}\\
\includegraphics[scale=0.3]{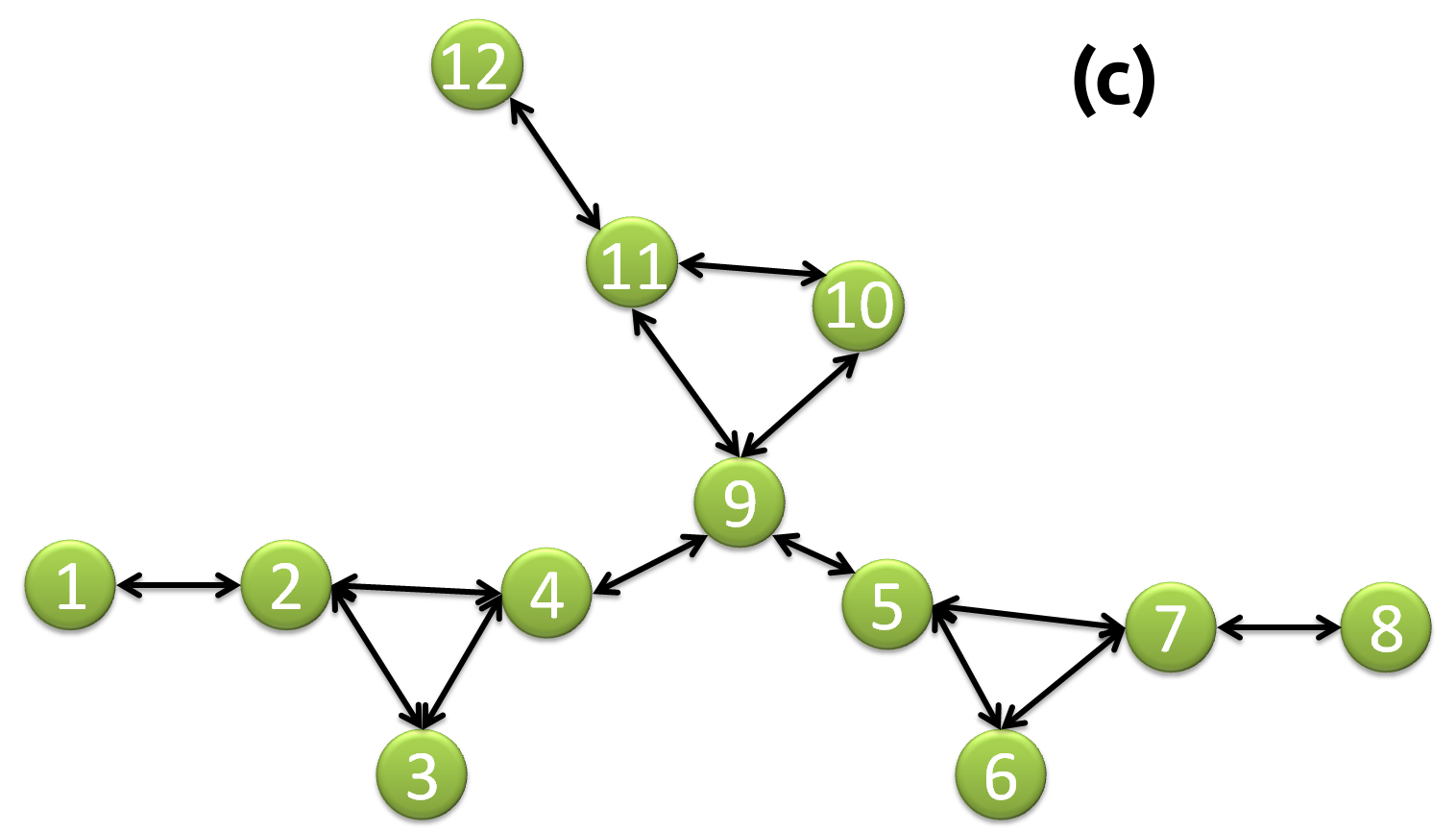}
\caption{Network topologies used {to construct {the} complex systems}. Panel {\bf (a)} shows a network with 16 nodes and {with similar characteristics} with a scale-free network, where the dynamics of each node {is} either a logistic or a circle map. Panel {\bf (b)} shows a network composed of 2 clusters of 3 nodes each, {which {is} composed of nodes with different time-scales and a logistic map dynamics}. Panel {\bf (c)} shows a network of 12 nodes, where the dynamics of each node is described by Hindmarsh-Rose dynamics [Eq.~\eqref{HR_model_1neuron}].}
\label{fig:networks}
\end{figure}

Figure~\ref{fig:networks}{\bf (a)} shows the network topology described by the adjacency matrix $\boldsymbol{A}_{ij}$ used to create a network where the dynamics of each node is described either by logistic or circle maps. We will use these networks to study the robustness of our methodology for different coupling strengths, observational noise, and data-length. We also use {small-size networks} with discrete dynamics, {with} different {decay of correlation times} for the nodes {to test our methodology} (see Fig.~\ref{fig:networks}{\bf (b)}). {In those networks,} the dynamics of the nodes is given by logistic maps. In particular, we construct a network formed by two clusters of 3 nodes each. The clusters are connected by a {small-coupling} strength link. {Specifically, the dynamics of} Fig.~\ref{fig:networks}{\bf (b)} for the cluster formed by the nodes $1, 2, 3$ is constructed by using $r=4$, and the dynamics of the cluster formed by the nodes $4, 5, 6$ is given by a third-order composition of the logistic map, i.e., $f(x^{i})\equiv f\circ f\circ f(x^{i})$, with $r=3.9$. Consequently, both clusters are constructed by time-series with different correlation {decay times}, creating a good example to understand how a {clustered network} with different time-scales can affect the inference capabilities of MI- or MIR-{based methodologies}.

\subsection{Networks with continuous-time units}
We consider continuous dynamics for the nodes of a network described by the Hindmarsh-Rose (HR) neuron model \cite{Hindmarsh-Rose_original_paper}. The {particular network we choose is shown} in Fig. \ref{fig:networks}{\bf (c)}. The HR model is given by
\begin{eqnarray}\label{HR_model_1neuron}
 \dot{p}&=&q-ap^3+bp^2-n+I_{\mbox{ext}}\nonumber,\\
 \dot{q}&=&c-dp^2-q\nonumber,\\
 \dot{n}&=&h[s(p-p_0)-n],
 \end{eqnarray}
where $p$ is the membrane potential, $q$ is associated with the fast currents ($Na^{+}$ or $K^{+}$), and $n$ with the slow current, for example $Ca^{2+}$. The rest of the parameters are defined as $a=1$, $b=3$, $c=1$, $d=5$, $s=4$, $p_0=-1.6$ and $I_{\mbox{ext}}=3.25$, for which the system exhibits a multi-scale chaotic behaviour with spike bursting. Parameter $h=0.005$ modulates the slow dynamics of the system. The neural networks of $M$ neurons connected by electrical (linear coupling) synapses is described in \cite{Hindmarsh-Rose_Baptista,Antonopoulosetal2015}, and corresponds to having
\begin{eqnarray}\label{HR_model_N_neurons}
 \dot{p}_i&=&q_i-a p_i^3+bp_i^2-n_i+I_{\mbox{ext}}-g_l\sum_{j=1}^{M}\mathbf{C}_{ij}H(p_j)\nonumber,\\
 \dot{q}_i&=&c-dp_i^2-q_i\nonumber,\\
 \dot{n}_i&=&h[s(p_i-p_0)-n_i],\;i=1,\ldots,M,
\end{eqnarray}
where $M$ is the number of neurons {and} $H(p_i)=p_i$. In Eq.~\eqref{HR_model_N_neurons}, $g_l$ is the strength of the electrical synapses. We use as initial conditions for each neuron $i$: $p_i=-1.30784489+\eta^r_i$, $q_i=-7.32183132+\eta^r_i$, $n_i=3.35299859+\eta^r_i$, and $\phi_i=0$, where $\eta^r_i$ is a uniformly distributed random number in $[0,0.5]$ for all $i=1,\ldots,N_n$, following \cite{Hindmarsh-Rose_Baptista,Antonopoulosetal2015}. $\boldsymbol{C}_{ij}$ is a Laplacian matrix and accounts for the way neurons are electrically (diffusively) coupled. Particularly, $\boldsymbol{C}_{ij} = \boldsymbol{K}_{ij}-\boldsymbol{A}_{ij}$ where $\boldsymbol{A}$ is the binary adjacency matrix of the electrical connections and $\boldsymbol{K}$ is the nodes degree diagonal matrix based on $\boldsymbol{A}$. If $\boldsymbol{A}(i,j)=1$ then neuron $j$ perturbs neuron $i$ with an intensity given by $g_l$.

\section{Methods} \label{sec:methodology}
\subsection{Calculation of the correlation {decay time} using the diameter of an itinerary network} \label{T_graph}

To infer the topology of a network using MIR [Eq.~\eqref{MIR_definition_epsilon}], we need to compute the correlation {decay time} $T(N)$. $T(N)$ is difficult to calculate in practical situations since it depends on quantities such as Lyapunov exponents and expansion rates, which demand a high computational cost \cite{Baptistaetal2012}. {Here, we} estimate it by the number of iterations that takes to points in cells of $\Omega$ to expand and completely cover $\Omega$. This is a necessary condition to determine the shortest time for the correlation to decay to zero. In particular, we are introducing a novel way to calculate $T(N)$ from the diameter of a network $G$, which is based on the dynamics of points mapped from one cell of $\Omega$ to another, namely, a network with the connectivity given by the transitions of points from cell to cell {of $\Omega$} or an itinerary network.

We construct $G$ as follows. We assume that each equally-sized cell in $\Omega$, occupied by at least one point, represents a node in $G$. Then, following the dynamics of points moving from one cell to another, we create the connections between nodes, i.e., the links in $G$. Specifically, a link between nodes $i$ and $j$ exists if points in $\Omega$ travel from cell $i$ to cell $j$. If the link exists the weight is equal to 1, if it is absent, then {it} is equal to $0$, therefore, $G$ is defined as a binary matrix with elements $G_{ij}\in\{0,1\}$. In this framework, a uniformly random time-series with no correlation results in a complete network, namely, an all-to-all network.
 
We define $T(N)$ as the diameter of $G$. The reason is that $T(N)$ is the minimum time that takes for points inside any cell of $\Omega$ to spread to the whole extent of $\Omega$. By definition, the diameter of a network is the maximum length for all shortest-paths, i.e, the minimum distance required to cross the entire network. Hence, our approach transforms the calculation of $T(N)$ into the calculation of the diameter of $G$. In particular, for the estimation of the network diameter we use Johnson's algorithm \cite{Johnson}. 

\subsection{Calculation of MIR}
{To estimate MIR from finite-length low-resolution time-series data, 
we truncate the summation in Eq.~\eqref{MIR_limits} up to a finite size $N$, depending on the resolution of data, and consider small trajectory pieces of the time-series with a length $L$, which depends on the total length of the time-series and on Eq. (\ref{noc}), such that,}
\begin{equation}
\mbox{MIR}_{XY} \cong \frac{1}{L}\sum_{i=1}^L[I_{XY}(i+1,N)-I_{XY}(i,N)].
\label{truncate}
\end{equation}
{In Eq. (\ref{truncate}), left-hand and right-hand sides would be equal if the partition, where probabilities are being calculated, is Markov.
The length $L$ represents also the largest order $T$ that a partition that generates statistically significant probabilities can be constructed from these many trajectory pieces. Assuming that the order of the partition constructed is $T=L$ (which also represents the time for the correlation in the partition to decay to zero, if the partition would be Markov), then Eq. (\ref{truncate}) becomes} 
\begin{equation}\label{eq:MIR_TIME}
\mbox{MIR}_{XY}\cong\frac{1}{T}\sum_{i=1}^T[I_{XY}(i+1,N)-I_{XY}(i,N)].
\end{equation}
Now, taking two partitions, $\Lambda_1$ and $\Lambda_2$, with different correlation {decay times}, $T_1$ and $T_2$ {respectively}, and different number of cells, $N_1 \times N_1$ and $N_2 \times N_2$ {respectively}, with $N_2>N_1$, {we have} $T_2=T_1+1$. Moreover, $\Lambda_1$ generates $\Lambda_2$ in the sense that $F^{-1}(\Lambda_1)=\Lambda_2$, where $F$ is the evolution operator and $F^{-1}(\Lambda_1)$ means the pre-iteration of partition $\Lambda_1$.
 Then, 
\begin{equation}
I_{XY}(T_2,\Lambda_1)=I_{XY}(T_1,\Lambda_2).
\end{equation}
Hence, we can write Eq.~\eqref{eq:MIR_TIME} as,
\begin{eqnarray} \nonumber
\mbox{MIR}_{XY}&\cong& \frac{1}{T_1}\sum_{i=1}^{T_1} [I_{XY}(i+1,\Lambda_1)-I_{XY}(i,\Lambda_1)]\vspace{0.25cm}\\
&\cong&\frac{1}{T_1}\sum_{i=1}^{T_1}[I_{XY}(i,\Lambda_2)-I_{XY}(i,\Lambda_1)].
\label{eq:MIR_partitions}
\end{eqnarray}

{When the partition is a} Markov generating partition, its properties  \cite{Baptistaetal2012} fulfil
\begin{equation}
I_{XY}(i,\Lambda_k)=I_{XY}(1,\Lambda_{k+i-1}).
\label{relate-markov}
\end{equation}
Then, {if our partition is close to a Markov partition}, Eq.~\eqref{eq:MIR_TIME} results in
\begin{eqnarray}\label{eq:MIR_markov_p}
\mbox{MIR}_{XY}&\simeq&\frac{1}{T_1}[I_{XY}(1,\Lambda_{T_1+1})-I_{XY}(1,\Lambda_1)]\\
&\equiv& \frac{1}{T_1}I_{XY}(1,\Lambda_{T_1}),
\end{eqnarray}
which is our demonstration for the validity of Eq.~\eqref{MIR_definition_epsilon}. 

{Therefore, in order to use Eq. (\ref{eq:MIR_markov_p}), we must have partitions for which 
Eq. (\ref{relate-markov}) is approximately valid. This condition can be reached {for} partitions constructed with a sufficiently large number of equally-sized cells of length $\epsilon=1/N$, exactly the type of partition considered here. Notice, however, that partitions will typically not be Markov {nor} generating, causing systematic errors in the estimation of MIR. To correct these errors, we propose the normalizations in Eqs. (\ref{MIR_epsilon_normalized}) and (\ref{MIR_en2}).}  

It is important to notice that $\mbox{MIR}_{XY}$ is always a partition-independent quantity, if and only if, the partitions are Markov. In order to calculate $I_{XY}(1,\Lambda_{T_1})$, we use Eq.~\eqref{eq:Shannon_MI}, which requires the calculation of probabilities in $\Omega$. Fulfilling the inequality
\begin{equation}\label{noc}
\langle N_{0}(N)\rangle \geq N_{oc},
\end{equation}
where $\langle N_{0}(N_{min})\rangle$ is the mean number of points inside all occupied cells of the partition of $\Omega$, Eq.~\eqref{noc} guarantees that the probabilities are unbiased.
 
\subsection{Network Inference using MIR}
For our analysis, using a non-Markovian partition allows us to simplify the calculations of $\mbox{MIR}_{XY}$, however, taking this kind of partitions {into consideration} would make the MIR values to oscillate around an expected value. Moreover, MIR for different non-Markovian partitions, not only has a non-trivial dependence with the number of cells in the partition, but also presents a systematic error \cite{steuer}. Therefore, since $\mbox{MIR}_{XY}$ for a non-Markovian partition of $N\times N$ equally-sized cells [estimated by Eq.~\eqref{MIR_definition_epsilon}], is expected to be partition-dependent, we propose here a way to obtain a measure, computed from $\mbox{MIR}_{XY}(N)$, that is {partition independent} and that is suitable for network inference.

To infer the structure of a network, we calculate the MIR for the $M(M-1)/2$ different pairs of nodes in the network, which is all we need due to the symmetric property of MIR. {We also} discard the MIR values for the same variable, i.e., {MIR$_{XX}$}, because we are interested in the exchange of information between different variables. We compute the $\mbox{MIR}_{XY}$ exchanged between any two nodes in a network by taking the expected value over different partition sizes $N_i$, i.e., $\mbox{MIR}_{XY}=E_i(\mbox{MIR}_{XY}(N_i))$, where $E(X)$ is the expected value of $X$. {In order to remove the systematic error \cite{steuer} in this calculation, we perform instead a} weighted average, where the finer partitions (larger $N$) contribute more to the $\mbox{MIR}_{XY}$ value than the coarser ones (smaller $N$). {The reason is that} a smaller $N$ is likely to create a partition that is further away from a Markovian one than a partition of larger $N$. Consequently, {we resolve the systematic error by weighing differently the different partitions}.

Therefore, we propose a novel normalization for the MIR as follows. First, we use an equally-sized grid of size $N$, we subtract from {MIR$_{XY}(N)$}, calculated for {all pairs of nodes,} its minimum value and denote the new quantity as  $\min(\mbox{MIR}_{XY}(N))$. Theoretically, a pair that is disconnected should have a MIR value close to zero, however, in practice, the situation is different because of the systematic errors coming from the use of a non-Markovian partition, as well as, from the information flow passing through all the nodes in the network. For example, the effects of a perturbation in one single node will arrive to any other node in a finite amount of time. This subtraction is proposed to reduce these two undesired overestimations of MIR. After this step, we remain with MIR as a function of $N$. Normalizing then by $\max(\mbox{MIR}_{XY}(N))-\min(\mbox{MIR}_{XY}(N))$, where again the maximum and minimum are taken over all different pairs, we construct a relative magnitude $\hat{\mbox{MIR}}_{XY}(N)$, namely,
\begin{equation}\label{MIR_epsilon_normalized}
\hat{\mbox{MIR}}_{XY}(N)=\frac{\mbox{MIR}_{XY}(N)-\min
\{\mbox{MIR}_{XY}(N)\}}{\max\{\mbox{MIR}_{XY}(N)\}-\min\{\mbox{MIR}_{XY}{(N)}\}},
\end{equation}
where $\mbox{MIR}_{XY}(N)$ is the MIR between nodes $X$ and $Y$ and $\min\{\mbox{MIR}_{XY}(N)\}$ is the minimum with respect to the $M(M-1)/2$ pairs and $\max\{\mbox{MIR}_{XY}(N)\}$ is the maximum with respect to {all $M(M-1)/2$ pairs}. This magnitude is still a function of $N$, however, we can now perform an average over different values of $N$ without the systematic error.
     
Next, we apply Eq.~\eqref{MIR_epsilon_normalized} for different grids sizes $N_i,\;i=1,\ldots,m$ to obtain $\mbox{MIR}_{XY}(N_i)$, where $N_m$ is the maximum number of cells per axis, resulting in a grid of $N_m\times N_m$ cells, and fulfilling at the same time Eq.~\eqref{noc}. Then, similarly to the idea used for Eq.~\eqref{MIR_epsilon_normalized}, we make a second normalization over $\hat{\mbox{MIR}}_{XY}(N_i)$ to obtain
\begin{equation}\label{MIR_en2}
\overline{{\mbox{MIR}}}_{XY}=\frac{\sum_{i}\hat{\mbox{MIR}}_{XY}(N_i)}{\max\{\sum_{i}\hat{\mbox{MIR}}_{XY}(N_i)\}},
\end{equation}
{where} the maximum {is} being taken now over the $N_m$ grids.

Finally, applying Eq.~\eqref{MIR_en2} to each pair ${XY}$, we obtain its average value, $\overline{{\mbox{MIR}}}_{XY}$. The higher the value of $\overline{{\mbox{MIR}}}_{XY}$, the higher the amount of information exchanged between $X$ and $Y$ per {unit of time}. This allows us to identify pairs of nodes that exchange larger rates of information than others. 

{In order to perform the network inference from the MIR}, we fix a threshold in $[0,1]$ and create a binary adjacency matrix $\boldsymbol{A}^c$, where the entry $\boldsymbol{A}^c_{X,Y}$ is $1$ if $\overline{\mbox{MIR}}_{XY}$ is higher than the threshold, and $0$ otherwise. $\boldsymbol{A}^c$ is then compared with the adjacency matrix $\boldsymbol{A}$ used to construct the dynamics of the nodes in Sec.~\ref{sec:section_models}. Recording the threshold used to create $\boldsymbol{A}^c$, and varying it in $[0,1]$, we obtain {different inferred networks. Our results show that there is an interval of thresholds within $[0,1]$} that fulfil $\boldsymbol{A}^c=\boldsymbol{A}$, i.e., a band that represents a $100\%$ successful network inference.

{In general, the effectiveness of our network inference {methodology} is measured by the absolute difference between the real topology and the one inferred for different threshold values. We find that whenever there is {a} band of threshold values, there is successful inference without errors. In practical situations, where the underlying network is unknown and the absolute difference is impossible to compute, the ordered values of the MIR or other similarity measures \cite{todos,Tirabassi2015} show a plateau which corresponds to the band of thresholds aforementioned. In particular, if the plateau is small, the authors in Ref.~\cite{Barabasi2013} propose a method to increase the size of the plateau by ``silencing'' the indirect connections, hence, allowing for a more robust reconstruction of the underlying network}.

\section{Results for Network Inference} \label{sec:results}
We now present our results for network inference using the three models introduced in Sec.~\ref{sec:section_models}.

 \subsection{Discrete-time systems}
\subsubsection{Different Coupling Strengths} \label{subsubsection_different_coupling_strengths}
Here we study the performance of Eq.~\eqref{MIR_en2} for network inference in the case where the dynamics of each node is described by a circle or a logistic map. The network structure that {comprises} our small-network of interacting discrete-time systems is given in Fig.~\ref{fig:networks}{\bf (a)}. {Here, we analyze the {effectiveness} of the inference} as the coupling strength, {$\alpha$,} between connected nodes, is varied. In Ref.~\cite{todos}, the authors have shown that, for the logistic [Eq.~\eqref{logistic_map_eq}] and circle maps [Eq.~\eqref{circle_map_eq}], assuming the same topology, the dynamics is quasi-periodic for $\alpha>0.15$ and chaotic for $0\leq\alpha\leq 0.15$. We, therefore, choose the coupling strength $\alpha$ in Eq.~\eqref{kaneko} to be equal to $0.03$ and $0.12$, both values corresponding to {chaotic dynamics}.

\begin{figure*}[htbp]
\centering
\includegraphics[scale=0.26]{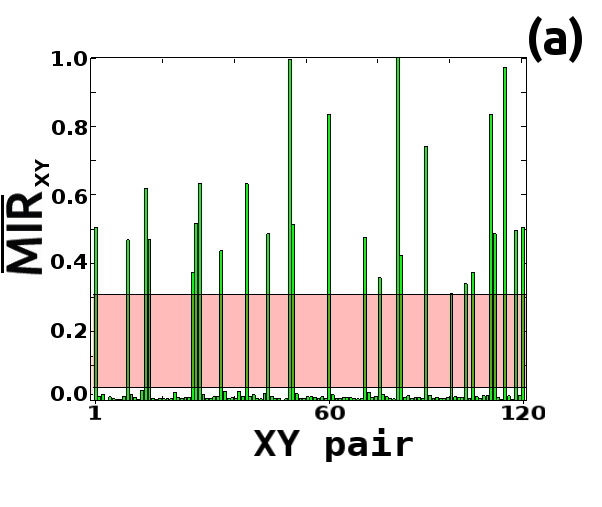}
\includegraphics[scale=0.26]{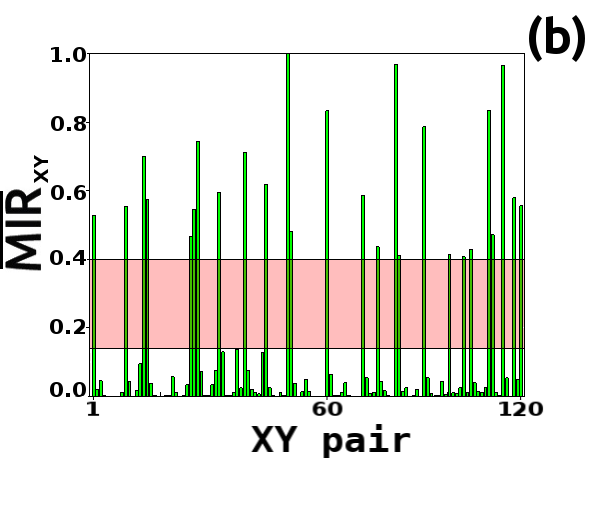}
\includegraphics[scale=0.26]{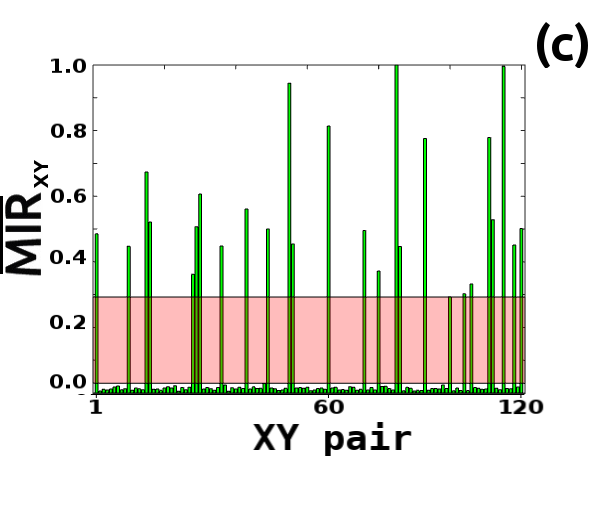}
\includegraphics[scale=0.26]{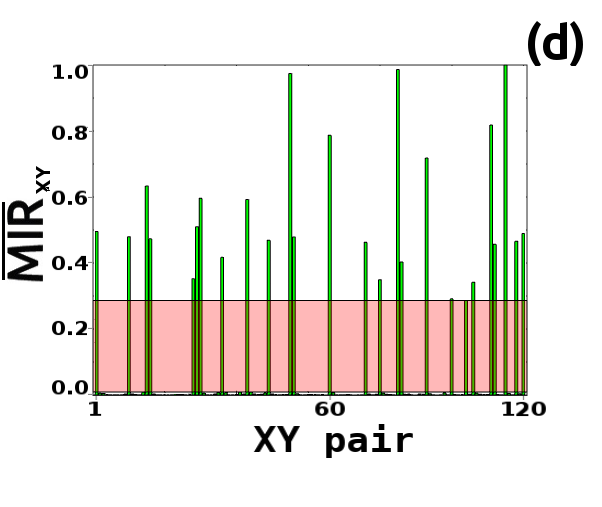}
\caption{Network inference for different coupling strengths and coupled maps. Panels {\bf (a)} and {\bf (b)} represent the $\overline{{\mbox{MIR}}}_{XY}$ values between different pair of nodes in a network composed {of} coupled logistic maps with a coupling strengths $\epsilon=0.03$ and $0.12$, respectively. Panels {\bf (c)} and {\bf (d)} are similar to panels {\bf (a)} and {\bf (b)}, but for circle maps. The red band indicates the range of thresholds from which the original network is correctly inferred, namely, {achieving 100\% successful inference}.}
\label{fig:coupling_strength_16_logistic_maps}
\end{figure*}

Figure~\ref{fig:coupling_strength_16_logistic_maps} shows the network inference results using $\overline{{\mbox{MIR}}}_{XY}$. The wideness of the red band represents all possible values a threshold can take to perform a 100\% success network inference, i.e. the correct identification of all physical and non-physical links. The wider the band, the bigger the probability to perform a complete reconstruction, therefore the reconstruction is more robust. When we deal with experimental data, and the correct topology is unknown, the optimal threshold can be determined by the range of consecutive thresholds for which the inferred topology is invariant, see Ref.~\cite{todos}.  

An error in the percentage of reconstruction comes from links that were not inferred (false negatives) or inferred erroneously (false positives). In our current study we avoid the distinction between them and we categorize both as reconstruction errors. Then, the reconstruction percentage can decrease by inferring non-existent links (non physical links) or by missing them. Each time this happens, we decrease the percentage by an amount $e_\%=100\frac{1}{N_l}$, where $N_l$ is the number of real links in the original network.

\subsubsection{Different time-series lengths and noise strengths} \label{subsubsection_different_time_series_lengths}
\begin{figure}[htbp]
\begin{center}
\includegraphics[scale=0.4]{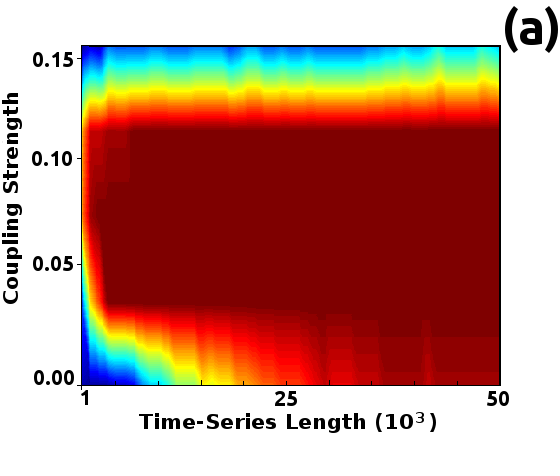}
\includegraphics[scale=0.4]{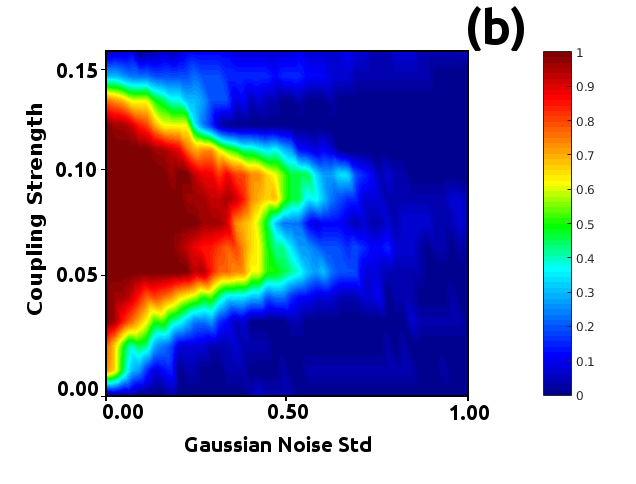}
\end{center}
\caption{Network inference based on  logistic maps, for different coupling and noise strengths. Panel {\bf (a)} shows the parameter space of the percentage of reconstruction (0\% - blue, 100\% - dark red) for different coupling strengths versus data lengths. Panel {\bf (b)} is similar to panel {\bf (a)} for different coupling strengths versus  the standard deviation of the normal distribution of the noise added.}
\label{fig:param}
\end{figure}

We start by analysing the effectiveness of $\overline{{\mbox{MIR}}}_{XY}$ for different time-series lengths, using the dynamics of the logistic map for each node and a coupling strength $\alpha\in[0,0.17]$. In Fig.~\ref{fig:param}{\bf (a)}, we observe that for $\alpha$ closer to $0.15$, a relatively short length (of about $3000$ points) is enough to infer correctly the original network, which is generated by the adjacency matrix {$\boldsymbol{A}$} of Sec.~\ref{sec:section_models}. However, when $\alpha$ is close to $0.03$, a larger time-series (of about $30000$ points) is needed to achieve $100\%$ successful reconstruction. Values of $\alpha=0$  and $\alpha\in[0.15,0.17]$ are considered to test the effectiveness of the values $\overline{{\mbox{MIR}}}_{XY}$ in the case of nodes being totally independent and in the case of nodes having periodic dynamics. In these regions, $\overline{{\mbox{MIR}}}_{XY}$ is expected to be zero, a situation evidenced in both panels of the figure. Our results so far suggest that the successful reconstruction for short-length time-series depends on the intensity of the coupling strength. However, it is surprising to see that exact inference can always be achieved for this dynamical regime if a sufficiently large time-series is available.

Next, we apply our methodology for network inference considering noisy time-series data. In particular, we introduce additive normally-distributed noise to the logistic map, i.e.,
\begin{equation}
f_{\mbox{ns}}(x_{n},r)=f(x_{n},r)+\gamma \cdot\sigma,
\end{equation}
where $f_{\mbox{ns}}(x_{n},r)$ is the noisy dynamics, $\sigma$ is a random number drawn from {the} normal distribution with $0$ {mean} and standard deviation of $1$, i.e. $\aleph(0,1)$, and $\gamma\in[0,1]$ is the noise strength. Since $\aleph(\gamma,1)=\gamma*\aleph(0,1)$, the noise strength {is} the standard deviation in the normal distribution. Fig.~\ref{fig:param}{\bf (b)} shows the parameter space for different coupling strengths versus $\gamma$. We observe perfect inference for noise strengths $\gamma<0.3$, i.e. for $\aleph(0,1)$. Moreover, the best reconstruction using $\overline{{\mbox{MIR}}}_{XY}$ is for coupling strengths in $[0.6,0.11]$, a dynamical regime where chaotic behaviour is prevalent.

\subsection{Neural Networks} \label{subsection_network_inference_from_neural_dynamics}
\begin{figure}[htbp]
\centering
\includegraphics[scale=0.26]{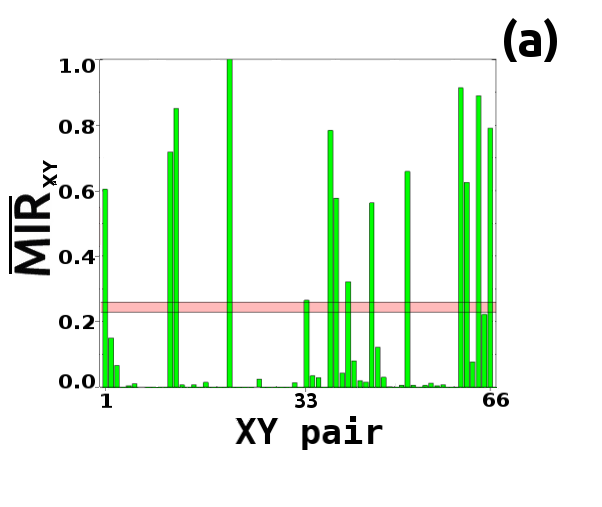}
\includegraphics[scale=0.26]{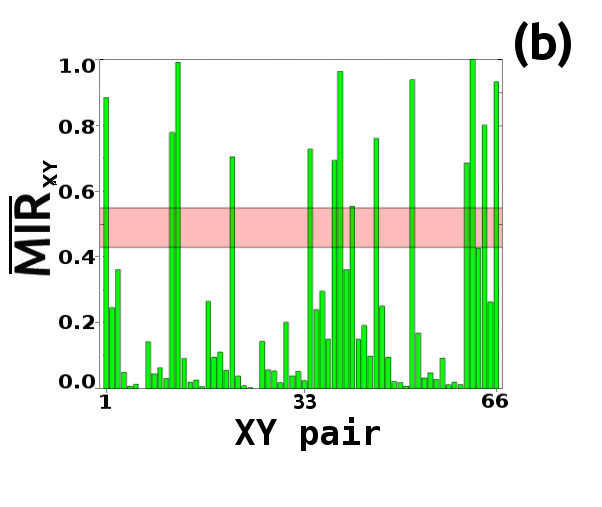}
\caption{Network inference for a network of nodes with a HR neural dynamics for different electrical couplings. Panels {\bf (a)} and {\bf (b)} show the bar plots of the percentage of inference for $g_l=0.05$ and $g_l=0.1$, respectively. The red bands show the range of thresholds for which the original network is inferred with a 100\% success.} 
\label{fig:Hindmarsh-Rose_recons}
\end{figure}

We also apply our methodology for the study of network inference in the case of continuous dynamics given by the HR system. We use two electrical couplings, $g_l=0.05$ and $0.1$, both considered for time-series of length $2\times10^5$. Figure~\ref{fig:Hindmarsh-Rose_recons} shows the band for 100\% successful network inference, where panel {\bf (a)} corresponds to $g_l=0.05$ and panel {\bf (b)} to $g_l=0.1$. This figure shows that $\overline{{\mbox{MIR}}}_{XY}$ is able to infer the correct network structure, in this case, for {small networks} of continuous-time interacting components.

\subsection{Comparison between Mutual Information and Mutual Information Rate} \label{subsection_MI_vs_MIR}
Finally, we compare MI and $\overline{{\mbox{MIR}}}_{XY}$ to assess the {effectiveness} of our proposed methodology for network inference. We apply the same normalization process used for MIR, Eq.~\eqref{MIR_en2}, to MI to have an appropriate comparison. In particular, we infer the network structure of the system described in Sec.~\ref{sec:section_models} with the network shown in Fig.~\ref{fig:networks}{\bf (b)}. As we have explained in Sec.~\ref{sec:section_models}, this system has two clusters of nodes with different dynamics. The dynamics in the left cluster is given by the 3rd-order composition of the logistic map, whereas the dynamics of the right cluster is given by ordinary logistic map dynamics. The different dynamics of the two groups produces different correlation {decay times}, $T(N)$, for nodes $X$ and $Y$, in particular when the pair of nodes comes from different clusters. The different correlation {decay times} produce a non-trivial dynamical behaviour that challenges the MI performance for network inference.
 
\begin{figure}[htbp]
\begin{center}
\includegraphics[scale=.90]{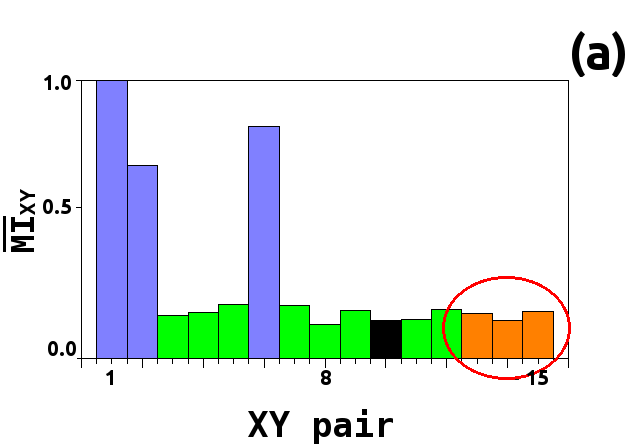}
\includegraphics[scale=.42]{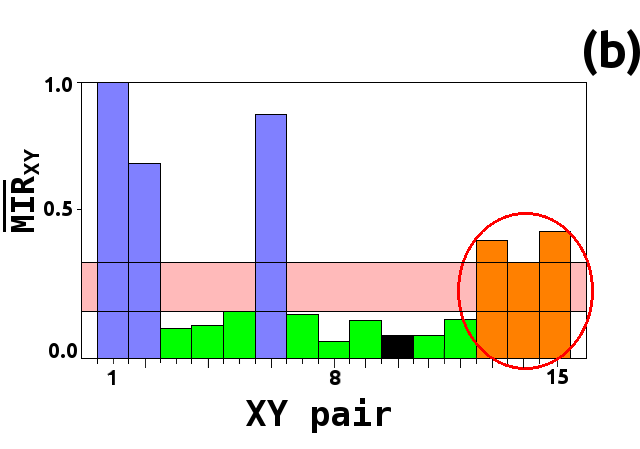}
\end{center}
\caption{Network inference for the composed dynamical system introduced in Sec.~\ref{sec:discrete_models}. Panel {\bf (a)} plots $I_{XY}$ of Eq.~\eqref{eq:Shannon_MI} for all links. This is a case where a complete network inference can not be achieved (indicated by the absence of any red band). Panel {\bf (b)} is the same as before but for $\mbox{MIR}_{XY}$. {The {color} code corresponds to the same {color} code identifying different nodes in Fig.~\ref{fig:networks}{\bf (b)}. The darkest {color} is the link connecting the two clusters.}}
\label{fig:composed_system}
\end{figure}

Figure~\ref{fig:composed_system} shows {the results obtained for the} normalized MI, $\overline{{I}}_{XY}$, and our normalized MIR, $\overline{{\mbox{MIR}}}_{XY}$, for each of the possible pairs of nodes. The purple bars correspond to the pairs of nodes $1$, $2$ and $3$ of the first cluster, the orange bars correspond to the pairs of nodes $4$, $5$ and $6$ of the second cluster (3rd order composed dynamics) and the black bar corresponds to the link between clusters (notice that due to the small coupling strength between the two clusters this link is not detected using any of the two methods). Nevertheless, MIR identifies correctly all intra links of the network where MI fails to do so. We conclude that the normalized MIR is preferable over the normalized MI when it comes to the detection of links in a complex system with different correlation {decay times}. The {reason is that the} normalized MIR takes into consideration the correlation {decay time} associated to each pair of nodes, {contrary to the MI}.

\section{Conclusions}\label{sec:conclusions}
In this paper we have introduced a new information based approach to infer the network structure of complex systems. MIR is an information measure that computes the information transferred per unit of time between pairs of components in a complex system. $\overline{{\mbox{MIR}}}_{XY}$, {our novel normalization for the MIR that is introduced} in Eq.~\eqref{MIR_epsilon_normalized}, is a measure based on MIR and developed for network inference. {We find that} $\overline{{\mbox{MIR}}}_{XY}$ is a robust measure to perform network inference in the presence of additive noise, short time-series, and also for systems with different coupling strengths. Since MIR and $\overline{{\mbox{MIR}}}_{XY}$ depend on the correlation {decay time} $T$, they are {suitable for inferring} the correct topology of networks with different time-scales.

{In particular, we have explored the effectiveness of MIR versus MI in terms of how successful they are in inferring exactly the network of our {small} complex systems. In general, we find that MIR outperforms MI when different time-scales are present in the system. {Our results} also show that both measures are sufficiently robust and reliable to infer the networks {analyzed} whenever a single time-scale is present. In other words, small variations in the dynamical parameters, time-series length, noise intensity, or topology structure, maintain a successful inference for both methods. It remains to be seen the types of errors that are found in these measures when perfect inference is missing or impossible to {be done}.}

\section{Acknowledgements}
EBM, MSB and CGA acknowledge financial support provided by the EPSRC ``EP/I032606/1'' grant. CGA contributed to this work while working at the University of Aberdeen and then, while {working} at the University of Essex, United Kingdom. NR acknowledges the support of PEDECIBA, Uruguay.

\newpage


\end{document}